\documentclass[a4paper, 12pt, final]{article}

\usepackage{amsmath, amssymb}
\usepackage{natbib}
\usepackage{fixltx2e}
\usepackage{mathtools}
\usepackage{amsfonts}
\usepackage{graphicx}
\usepackage{xfrac}
\usepackage{topcapt}
\usepackage{array}
\usepackage{tabularx}
\usepackage{setspace}
\usepackage{float}
\usepackage{tabularx}
\usepackage{lineno}
\usepackage[top=2.4cm, bottom=2.4cm, left=2.4cm, right=2.4cm]{geometry}
\floatstyle{boxed}
\restylefloat{figure}
\DeclareGraphicsExtensions{.pdf,.png,.jpg}

\title{Examining posterior propriety in the Bayesian analysis of capture-recapture models}
\author{Arjun M. Gopalaswamy\textsuperscript{*1,2} and Mohan Delampady\textsuperscript{1}}

\begin{document}
\maketitle
\emph{\textsuperscript{1}Statistics and Mathematics Unit, Indian Statistical Institute, Bangalore Centre, Bengaluru, 560059, INDIA}\\
\emph{\textsuperscript{2}Department of Zoology, University of Oxford, South Parks Road, Oxford - OX1 3PS, UK}\\

\emph{\textsuperscript{*}Corresponding Author:} Arjun M. Gopalaswamy (arjungswamy@gmail.com)\\
\emph{Email addresses of other authors:} Mohan Delampady (mohan@isibang.ac.in)\\



\begin{spacing}{1.0}
\begin{abstract}
There lies a latent danger in utilizing some known mathematical results in ecology. Some results do not apply to the problem at hand. We identify one such trend. Based on a couple of theorems in mathematical statistics, Link (2013) cautions ecologists about the inappropriateness of using the discrete uniform prior in their analysis under certain conditions and instead recommends the routine use of the scale prior during analysis. This recommendation is been absorbed immediately and widely among ecologists. In this study, we consider the two fundamental capture-recapture models used widely in ecology, $M_0$ and $M_h$, and derive conditions for posterior propriety by examining the behavior of the right tail of the posterior distributions of animal population size $N$ in a Bayesian analysis. We demonstrate that both these likelihoods are far more efficient than the ones considered in Link (2013). We argue that no particularly prescriptive approach should be adopted by ecologists in regard to choosing priors of the fear of posterior impropriety. Instead, we recommend the efficient construction of likelihoods for the problem and data on hand, choosing priors based existing knowledge of a parameter of interest and encourage examining posterior propriety by asymptotic arguments as demonstrated in this study. 
\end{abstract}

\section{Introduction}

Mathematics has played a crucial role in the development of concepts in ecology over the past 50 years. However, as noted earlier by \citet{may2004uses}, there lies a latent danger in utilizing some known mathematical results in the life sciences. This can happen because as a closed form of study, the rules defined by mathematics lead to impenetrable results. The tightness of these results can sometimes impart a fear on practicing biologists if the applicability of such mathematical results are not critically assessed against the problem at hand. As a result some of these mathematical results may sometimes mislead, rather than lead, empiricists, and is not desirable. Here, we identify one such instance. 

In this study we take up a specific example in statistical ecology of a recently published study by \citet{link2013cautionary}. In this study, \citet{link2013cautionary} discusses certain oddities of using the discrete uniform prior ($\propto 1$) in the Bayesian data analysis of the estimation of animal abundance $N$. \citet{link2013cautionary} discusses how, under certain conditions, the discrete uniform prior yields an improper posterior distribution of $N$ and also shows how this prior can lead to ``Bayesian stupefaction'' - an counter-intuitive process where an increase in data or samples leads to poorer knowledge.  Based on these oddities, \citet{link2013cautionary} recommends ecologists to \emph{routinely} use the scale prior ($\propto 1/N$) over the discrete uniform prior ($\propto 1$) in the Bayesian analysis of such abundance estimation problems, to avoid these oddities. 

Following this recommendation, several studies henceforth have immediately adopted new practices. To name a few: \citet{augustine2014accounting} have utilised this recommendation in attempting to find ways of removing biases associated with behavioural responses in hair snare mark-recapture surveys, \citet{bonner2014mc} utilise this recommendation while exploring the use of Monte Carlo integration within MCMC in mark-recapture problems, \citet{broms2015accounting} utilise it in estimating Hill numbers for biodiversity studies, \citet{conn2015using} whets the use of the scale prior in constructing spatiotemporal statistical models for abundance estimation cautioning against impropriety of the posteriors, and, \citet{gerber2015spatial} use the recommendation in evaluating spatial capture-recapture estimators in known small mammal densities. We worry, however, whether all of these studies have critically assessed whether the findings of \citet{link2013cautionary} apply to their particular problems or not.   

The basis for the recommendations made in \citet{link2013cautionary} lay in theorems developed in two mathematical papers describing certain features of specific versions of the binomial \citep[see][]{kahn1987cautionary} and multinomial \citep[see][]{york1992bayesian} likelihoods, respectively. The mathematical results from these studies, specifically about conditions on the prior causing improper posterior distributions, form the basis for the conclusions of \citet{link2013cautionary}. Impropriety of a posterior distribution means that the posterior distribution does not integrate to one and this causes problems in interpreting posteriors. However, the likelihoods used in \citet{kahn1987cautionary} and \citet{york1992bayesian} are rarely used in the abundance estimation problems in ecology. Hence, we are concerned that the recommendation made by \citet{link2013cautionary} may not be relevant to some of the applications discussed above. 

Here we take up two basic capture-recapture models that are used widely in ecology, $M_0$ and $M_h$ \citep{otis1978statistical}, and assess, independently, the particular concern of posterior propriety of the parameter of central interest to us, animal abundance $N$. In specific, we:
\begin{enumerate}
\item Compare and contrast the \citet{kahn1987cautionary} likelihood with the $M_0$ likelihood. 
\item Compare and contrast the \citet{york1992bayesian} likelihood with a specific version of the $M_h$ model likelihood.   
\item Derive two novel theorems of propriety for the $M_0$ and $M_h$ likelihoods, respectively, by examining the asymptotic (when $N$ is large) behavior of these functions. For such an examination, we employ the Big-$O$ notation, that is widely used in computer science and complexity theory \citep{knuth1976big}. 
\item Discuss the broader relevance of these theorems to practicing ecologists.  
\end{enumerate}

\section{Methods}

The arguments of propriety posed by \citet{link2013cautionary} are based on two fundamental theorems derived in \citet{kahn1987cautionary} and \citet{york1992bayesian}, respectively. In essence, the conditions of propriety discussed in \citet{kahn1987cautionary}, which is based on a version of the binomial model, find simple extensions to \citet{york1992bayesian}, which is based on a version of the multinomial model. Rather than going into the details and investigating particular aspects of \citet{link2013cautionary} note, we \emph{directly} consider the models utilized in \citet{kahn1987cautionary} and \citet{york1992bayesian}, respectively, discuss their relevance to ecology (particularly in light of models developed in the animal abundance estimation literature) and develop our arguments from thereon. 

\subsection{Comparing models of \citet{kahn1987cautionary} and the classical $M_0$ model}

Let $N$ be the total number of animals in the study that we are interested to estimate and let $p$ be the capture probability of each of these animals on any of $K$ sampling occasions. In the development of \citet{kahn1987cautionary} we have, $N_1, N_2, \dots N_K$ as iid binomial $(N,p)$ random variables. If the observations are $n_1, n_2, \dots  n_K$, then the likelihood is 
\begin{equation}
L(N,p;\{n_j\}) \propto p^{[\sum_{j=1}^{K}n_j]} \times (1-p)^{[KN-\sum_{j=1}^{K}n_j]} \times
 \prod_{j=1}^{K} \frac{N!}{(N-n_j)!}, 
\end{equation}
since the probability of the observations $n_1, n_2, \dots  n_K$ is given by
\begin{eqnarray}
P(n_1, n_2, \ldots ,n_K|N,p) &=&  p^{[\sum_{j=1}^{K}n_j]} \times (1-p)^{[KN-\sum_{j=1}^{K}n_j]} \times
 \prod_{j=1}^{K} \frac{N!}{n_j! (N-n_j)!} 
\nonumber\\
&=& p^{n_{.}} \times (1-p)^{[KN-n_{.}]} \times \prod_{j=1}^{K} \frac{N!}{n_j! (N-n_j)!}. \label{kahn-model}
\end{eqnarray}
where $n_{.}=\sum_{j=1}^{K}n_j$ and $n_j$ is the number of captures on occasion $j$. 

Now consider the capture-recapture $M_0$ model where again we have only two parameters, $p$ and $N$ just as above. Let
$$y_{ij} = \left\{\begin{array}{cl}
1 & \mbox{ if animal $i$ is captured on occasion $j$;}\\
0 & \mbox{ otherwise.}
\end{array}\right.$$
Here $\{y_{ij}\}$ ($1\le i\le N$, $1\le j\le K$) is the capture-recapture history of all the animals. We further assume that $M_{K+1}$ distinct animals are captured at least once, so that $N-M_{K+1}$ animals
exhibit a string of 0s as their capture histories. The joint probability distribution of the entire capture-recapture history takes the form 
\begin{eqnarray}
P(\{y_{ij}\}|N,p) &=& {N\choose M_{K+1}} (1-p)^{K(N-M_{K+1})}\prod_{i=1}^{M_{K+1}}\prod_{j=1}^K p^{y_{ij}}(1-p)^{1-y_{ij}}
\nonumber\\
&=& \frac{N!}{[N-M_{K+1}]![M_{K+1}]!}(1-p)^{K(N-M_{K+1})}\prod_{j=1}^K p^{n_j}(1-p)^{M_{K+1}-n_j} \nonumber\\
&=& \frac{N!}{[N-M_{K+1}]![M_{K+1}]!} p^{n_{.}}(1-p)^{KN-n_{.}},\label{cr-model}
\end{eqnarray}
where $n_{.}=\sum_{j=1}^{K}n_j$ and $n_j = \sum_{i=1}^{M_{K+1}} y_{ij}$ is the number of captures on occasion $j$.

From the two likelihoods (\ref{kahn-model} and \ref{cr-model} estimating $(N,p)$) we observe that the powers associated with $p$ and $1-p$  between the two models is exactly the same. So, qualitatively, $\hat{p}$ and $\hat{N}$ can be estimated in the same way. That is, by substituting the expression $\hat{p} = n_{.}/KN$, which is just the number of captures divided by the number of opportunities for capture, into the likelihood function and simply maximizing with respect to $N$. This yields the MLE $\hat{N}$ for population size - the parameter we are mainly interested in. 

However, there exists a difference in the structure of the combinatorial terms. In the $M_0$ model  of \citet{otis1978statistical}, we are paying particular attention to the total number of individuals observed in the study $(M_{K+1})$, which is ignored in (\ref{kahn-model}) of the \citet{kahn1987cautionary} model. The idea is that as $K$ increases, $M_{K+1}$ also increases and this will continue until $M_{K+1} = N$ after which the increase in $K$ will not matter.  With the same idea, the cumulative probability, $P=1-(1-p)^K$ which is the probability of an individual being caught at least once, increases as $K$ increases, leading ultimately to the situation that when $P=1$, $M_{K+1} = N$. This additional piece of information about the individuals captured $M_{K+1}$, is lost in (\ref{kahn-model}). 

Owing to such differences in the model structure itself, we may presume that the likelihood construction in \citet{kahn1987cautionary} simply does not efficiently make use of all the information that might be available in data as standard capture-recapture models \citep{otis1978statistical, williams2002analysis} do, or even for that matter, the basic $M_0$ model. Therefore, we may argue that when such an inefficient likelihood with parameter identifiability issues, such as (\ref{kahn-model}), is used, we may naturally expect a greater influence of the choice of the prior distribution in determining posteriors during a Bayesian analysis. 

\subsection{Comparing the conditions for propriety in the \citet{kahn1987cautionary} model relative to the $M_0$ model}

Both the models (\ref{kahn-model}) and (\ref{cr-model}) aim at estimating $N$ and $p$ from specifically obtained data. A Bayesian analysis of this problem will involve specifying priors for $N$ and $p$. The question we are interested in, and brought to light by \citet{link2013cautionary}, is when a particular prior model is specified for $p$, what should be the prior model we should specify for $N$, so that the posterior distribution for $N$ is proper. In particular, we are interested to find propriety conditions of the posterior, given a specific prior model on $p$ - for example, an \emph{a priori} belief of $Unif(0,1) \equiv Beta(1,1)$ model. More generally, we can simply define this as $\pi_1(p) = Beta(a,b)$, where $a$ and $b$ are the shape parameters of the beta distribution that will describe our \emph{a priori} belief around the parameter $p$ before the conduct of the experiment. 

From the theorem derived in \citet{kahn1987cautionary}, if we were to conduct a Bayesian analysis on model (\ref{kahn-model}) by imposing a prior structure of $\pi_1(p) = Beta(1,1)$, simultaneously, with a discrete uniform prior on $N$ ($\pi_2(N) \propto 1$), then we would obtain impropriety in the posterior distribution of $N$. Instead, if we choose the scale prior for $N$ ($\pi_2(N) \propto 1/N$), we obtain a proper posterior distribution for $N$ \citep{link2013cautionary}. Due to this particular situation, \citet{link2013cautionary} recommends ecologists to routinely utilise the scale prior over the discrete uniform prior to avoid such resulting impropriety in the posterior distributions. We therefore ask whether such a blanket rule should apply in all situations. We therefore begin by investigating the conditions for posterior propriety of $N$ on the $M_0$ model - one of the most basic abundance estimation models in the capture-recapture literature \citep{otis1978statistical}.  

\subsubsection{Establishing conditions of posterior propriety of $N$ for Model $M_0$}

\paragraph{Big-$O$ Notation} 

It is often useful to talk about the \emph{rate} at which some mathematical function changes as its argument grows (or shrinks), without worrying much about the detailed form. In mathematics, this is what the notations $O(.)$ lets us do. This notation enable us to assess the asymptotic behavior of functions in mathematics.  Since the evaluation of posterior propriety requires such assessments, we use the Big-$O$ notation for specific problems of interest to us here. There is also the little-$o$ notation, but we do not need it here. This broad approach is widely used in understanding the efficiency of computer algorithms  and in complexity theory \citep{knuth1976big}. For example, if we have a function $T(m)=5m^3+4m^2+m+12$, we can say that the function `$T(m)$ grows by an \emph{order} of $m^3$', since the first term dominates over the other terms asymptotically. We would write this down as $T(m)=O(m^3)$. 

In our problem here, we note that integrability of a density over nonnegative integers (and existence of its moments) will depend on its behavior in the right tail. That means, in essence, we are studying the asymptotic behavior of the posterior functions of $N$, when $N \to \infty$. We use the Big-$O$ notation here to indicate the rates of growth (or decline) of these functions as $N \to \infty$. For propriety, what we want is that as $N \to \infty$, the posterior function must decline, and at a rate faster than $O(1/N)$. Because $O(1/N)$ again indicates impropriety by the same argument that the scale prior ($\propto 1/N$) is improper \citep{link2013cautionary}. For this reason, when the posterior is $O(1/N^d)$, where $d>1$, the function implies propriety. Thus, we arrive at the following theorem for the $M_0$ model. 

\noindent {\bf Theorem 1: (for the $M_0$ model).} For the Bayesian analysis of model (\ref{cr-model}), consider the joint prior density $\pi(p,N) = \pi_1(p) \pi_2(N)$ where $\pi_1(p)$ is the density of
Beta($a$,$b$). Then the posterior distribution of $N$ is proper if and only if
$n_{.}-M_{K+1} +a > 1$ for the discrete uniform prior $\pi_2(N) \propto 1$; and it is
proper if and only if $n_{.}-M_{K+1} +a > 0$ for the scale prior, $\pi_2(N) \propto 1/N$.

\noindent {\bf Proof.} We prove the above theorem by showing that  $\pi\left(N|\{y_{ij}\}\right)=\pi_2(N)O\left(N^{-(n_{.}-M_{K+1}+a)}\right)$ for model (\ref{cr-model}). 

See Appendix (6.1) for the detailed proof of this Theorem.

\subsection{Comparing the conditions for propriety in the \citet{york1992bayesian} model relative to the $M_h$ model}

Drawing upon the same arguments provided above, we can similarly compare the likelihood of the \citet{york1992bayesian} model with the $M_h$ model used in capture-recapture literature \citet{otis1978statistical}. In simple terms, \citet{york1992bayesian} merely extend the arguments of \citet{kahn1987cautionary} from the binomial model to the multinomial model. Just as the beta distribution forms the natural choice of priors for the constant detection probability parameter $p$ in a binomial problem, the Dirichlet distribution forms the natural choice of a prior for the vector $\bf p$ in a multinomial distribution. 

Accordingly, \citet{york1992bayesian} arrive at two propositions for arriving at the propriety of the posterior of $N$ for given prior conditions. They show that when the prior $\pi(N) \propto 1$, then the posterior $\pi(N|D)$ is proper if and only if $\delta > 1/(k-1)$, where $k$ is the number of cells in the Dirichlet prior, $\delta$ is the parameter for each cell in the Dirichlet distribution and $D$ are data. They also show that when the prior $\pi(N) \propto 1/N$, then the posterior $\pi(N|D)$ is proper for any $\delta>0$. This result leads \citet{link2013cautionary} to conclude that the scale prior must be routinely used owing to its ability to impose fewer conditions for propriety. 

It does, therefore, become relevant to ask whether the \citet{york1992bayesian} models apply directly to abundance estimation models that we use often in ecology. With this motivation, we try and establish the comparison between a specific version of the $M_h$ capture-recapture model \citet{otis1978statistical}, that is widely used in ecology, and the \citet{york1992bayesian} model. 
 
The $M_h$ model is more complicated relative to the $M_0$ model. The model is parameterized with $N$ capture probabilities $p_1, \dots p_N$ as well as the population size, $N$, for a total of $N+1$ parameters. This leads to a large number of parameters, and as a result, initially, \citet{burnham1978estimation}  considered alternatives for ML estimation for this model. Here the approach was to treat the vector of capture probabilities $\{p_i\}$ as a random sample of size $N$ from some probability distribution $F(p)$ defined on the interval $[0,1]$. The corresponding statistical model can be described in terms of the number $f_j$ of animals caught on exactly $j$ occasions. 
 
\begin{equation}
P(f_1, \dots , f_K|F) = \frac{N!}{[\prod_{j=1}^{K}f_j!](N-M_{K+1})!} \pi_0^{N-M_{K+1}}\prod_{j=1}^{K}\pi_j^{f_j}  \label{mhsummary-model}
\end{equation}  

where, 
\[
\pi_j = \int_0^1 \frac{K!}{(K-j)!j!}p^j(1-p)^{K-j}dF(p).
\]

Here, the cell probability $\pi_j$ can be viewed as the average probability that an individual
is caught exactly $j$ times. And $F(p)$ can be any class of probability distributions, and naturally can be the beta distribution.
 
York and Madigan (1992) develop the posterior, after integrating out the probability measure $\theta$, as
\begin{eqnarray}
P(N|D) & \propto & P(D|N)P(N)\nonumber\\
& \propto & \frac{\Gamma(N+1)}{\Gamma(N-n+1)} \times \frac{\Gamma(N-n+\delta)}{\Gamma(N+k\delta)}P(N).\label{york-madigan}
\end{eqnarray}

As described earlier, $K$ is the number of cells of the Dirichlet prior ($K=2$ for the binomial case) and $\delta$ is the parameter for each cell of the Dirichlet distribution. And $n$ is the number of cases or successes actually observed. Once again of particular interest to us is the observation that (\ref{york-madigan}) does not contain an explicit structure on $p$, $F(p)$, that is present in (\ref{mhsummary-model}). Without this structure the problem will involve a lot of nuisance parameters. However, \citet{york1992bayesian} were motivated to develop a variety of ways to establish complex relationships between administrative lists which would eventually determine the number of parameters in the model. This complication analogously translates to investigating complex relationships amongst the $K$ sampling occasions in a standard capture-recapture sampling situation \citep{otis1978statistical}. This complication does not appear much in capture-recapture literature, but that is not to say it is not relevant in ecology. But, that is not the focus of our study here. Our aim, instead, is to encourage a careful, independent, evaluation of posterior propriety of parameters when specific statistical models are used by ecologists to solve their problems. In that context, we can intuitively expect that the $M_h$ models discussed in the capture-recapture literature \citep{williams2002analysis, amstrup2005handbook} may not demand stringent conditions to yield proper posteriors.  

\subsubsection{Establishing the posterior propriety conditions for the $M_h$ model}

Instead of utilizing the model (\ref{mhsummary-model}) that, in some ways, captures only summary information from the data, we will consider the complete data likelihood. The complete data likelihood is also adopted by others \citep[see][]{bonner2014mc, king2015capture} for different reasons.  

\begin{eqnarray}
P(\{y_{ij}\}|N,\{p\}) &=& \frac{N!}{[M_{K+1}]! [N-M_{K+1}]!}\prod_{i=1}^{M_{K+1}}p_i^{y_{i.}}(1-p_i)^{K-y_{i.}}
\nonumber\\
& & \times \left(\prod_{i=M_{K+1}+1}^N (1-p_i)^K\right),\label{mh-cr-model}
\end{eqnarray}
where $y_{i.} = \sum_{j=1}^K y_{ij}$.
As is normally done, we assume that $p_i$ are randomly drawn from a Beta($\alpha, \beta$)
population. Then (\ref{mh-cr-model}) leads to
\begin{eqnarray}
\lefteqn{P(\{y_{ij}\}|N,\alpha,\beta)}\nonumber\\
&=& \frac{N!}{[M_{K+1}]! [N-M_{K+1}]!}
\int_{[0,1]^N}\prod_{i=1}^{M_{K+1}} p_i^{y_{i.}}(1-p_i)^{K-y_{i.}}
 \left(\prod_{i=M_{K+1}+1}^N (1-p_i)^K\right)\nonumber\\
&& \times \prod_{i=1}^N \left(\frac{\Gamma(\alpha+\beta)}{\Gamma(\alpha)\Gamma(\beta)}
p_i^{\alpha-1}(1-p_i)^{\beta-1}\right)\,d\{p\}\nonumber\\
&=&  \frac{N!}{[M_{K+1}]![N-M_{K+1}]!} \prod_{i=1}^{N-M_{K+1}}
\frac{\Gamma(\alpha+\beta)}{\Gamma(\alpha)\Gamma(\beta)}
\frac{\Gamma(\alpha+y_{i.})\Gamma(\beta+K-y_{i.})}{\Gamma(\alpha+\beta+K)}\nonumber\\
&& \times \left(\prod_{i=M_{K+1}+1}^N \frac{\Gamma(\alpha+\beta)}{\Gamma(\alpha)\Gamma(\beta)}
\frac{\Gamma(\alpha)\Gamma(\beta+K)}{\Gamma(\alpha+\beta+K)}\right)\nonumber\\
&=&  \frac{N!}{[M_{K+1}]![N-M_{K+1}]!}
\left(\frac{\prod_{j=0}^{K-1}(\beta+j)}{\prod_{j=0}^{K-1}(\alpha+\beta+j)}\right)^{N-M_{K+1}}\nonumber\\
&&\times \prod_{i=1}^{M_{K+1}}\left(\frac{\prod_{j=0}^{y_{i.}-1}(\alpha+j)\prod_{j=0}^{K-y_{i.}-1}(\beta+j)}
{\prod_{j=0}^{K-1}(\alpha+\beta+j)}\right).
\label{int-mh-cr-model}
\end{eqnarray}

Thus, we can arrive at another theorem for the model (\ref{int-mh-cr-model}). 

\noindent {\bf Theorem 2. (for the $M_h$ model)} For the Bayesian analysis of model (\ref{int-mh-cr-model}), consider the joint prior density $\pi(\alpha,\beta ,N) = \pi_1(\alpha,\beta) \pi_2(N)$ where, under $\pi_1$, $\alpha$ and $\beta$ are independent Gamma random variables with shape parameters $a>0$ and $b>0$ respectively and a common scale parameter $c$. Then the posterior distribution of $N$ is proper if $a > 1$ for the discrete uniform prior $\pi_2(N) \propto 1$; and it is proper if $a > 0$ for the scale prior, $\pi_2(N) \propto 1/N$. 

\noindent {\bf Proof.} We prove the above theorem by showing that $\pi(N|\{y_{ij}\}) = \pi_2(N) O(N^{-a})$ for model (\ref{int-mh-cr-model}).  
See Appendix (6.2) for a detailed proof of this theorem. 

\section{Conclusions}

Let us impose a constraint that the prior for $p$ comes from a $Beta(1,1) \equiv Unif(0,1)$ because this forms the basis of the arguments discussed in \citet{link2013cautionary}. So, $a=1$. From the theorem for the $M_0$ model, we conclude that if the number of recaptures $r = n_{.}+M_{K+1}\geq 1$, then the posterior distribution for $N$ is proper when the prior distribution for $N$ is a discrete uniform prior ($\propto 1$). Similarly, the posterior for $N$ is proper when the prior distribution for $N$ is a scale prior ($\propto 1/N$) as long as the number of recaptures $r \geq 0$. The result should not be very surprising even to those who remain oblivious to issues concerning posterior propriety. Heuristically speaking, information for the parameter $p$ comes from some form of replication in capture-recapture experiments. Since, the prior for $p$ is already set to $Beta(1,1)$, all the information from the data for $p$ has to come from the recaptures $r$ of individuals in the study. Obviously, if we do not have any recaptures, when $r=0$, then we clearly have a problem of identifiability between $N$ and $p$ due to lack of information. Hence, when the discrete uniform prior is used for $N$, when prior for $p$ is $Beta(1,1)$ and $r=0$, the resulting posterior for $N$ is bound to be improper simply due to this identifiability issue. However, we can still ensure propriety when a scale prior is used for $N$ even when $r=0$. But, it is to be noted, however, that the choice of a scale prior for $N$ does not in any way imply that the posterior for $N$ is now any more informative than in the discrete prior choice for $N$, because information for $p$ has to still come from recaptures (to inform the likelihood) or from another source (which would change the prior for $p$). 

Similarly, the version of the $M_h$ model considered in this study involves drawing $p_i$ from a random $Beta(\alpha,\beta)$ population. Under the joint prior $\pi_1(\alpha, \beta)$, $\alpha$ and $\beta$ are independent Gamma random variables with shape parameters $a>0$ and $b>0$ with a common scale parameter. Here, the posterior distribution of $N$ is proper if $a>1$ for the discrete uniform prior $\pi_2(N) \propto 1$. This argument relates to the discussion made above for the $M_0$ case. However, it is difficult to describe this argument heuristically. What we can infer though is that when $a \leq 1$ for the discrete uniform case, we will observe a large accumulation of probability mass at values very close to 0, and very little heterogeneity in capture probability $\bf{p}$. And when there is little or no heterogeneity at all, and the capture rates are very low (owing to a large probability mass very close to 0), we may once again visualize a case of parameter redundancy during estimation. As with arguments made for the $M_0$ case above, by invoking the scale prior for $N$, a value of $a>0$ is sufficient enough to yield proper posteriors for $N$. But once again, posterior propriety of $N$ does not mean that the posterior is any more informative. It is interesting to note that $b$ does not appear in the condition, and indeed it might be possible to arrive at propriety conditions involving $b$. But we note that when $a$ increases, it implies that the average $p$ increases, and consequently informs us that $M_{K+1}$ approaches $N$. In that case, while the heterogeneity model may not be an efficient model for estimation, relative to the $M_0$ model, there is perhaps no specific concern of posterior impropriety.

\section{Discussion}

\citet{link2013cautionary} brings up two very important issues for practitioners of Bayesian data analysis. The study makes a persuasive case for the need of ensuring propriety in the posterior distribution of a parameter of ecological interest. The study also recognizes an important oddity that occurs when the choice of the prior sometimes provides illogical inference, particularly by demonstrating a case of how `stupefaction' can occur even while we accumulate more data in the scientific experiment.  We concede that both of these data analytical issues merit serious attention. 

We note that \citet{link2013cautionary} makes a recommendation that the scale prior ($\propto 1/N$) should be \emph{routinely} favoured over discrete uniform prior ($\propto 1$) owing to the fact that the discrete uniform prior can \emph{sometimes} lead to the oddities discussed above. However, more generally, we are concerned about any such blanket application of a rule in the practice of Bayesian inference. While the concerns raised by \citet{link2013cautionary} are exemplified by the choice of the binomial likelihood, combined with the choice of priors in the study, we argue that such likelihoods are rarely used in ecology, and, in fact, even some of the most basic likelihood constructions of statistical ecologists, say the $M_0$ or the $M_h$ models \citep{otis1978statistical}, are efficient enough that such problems of propriety do not appear. 

In this study, we make use of the Big-$O$ notation to examine the asymptotic behavior of the likelihood functions of the $M_0$, and a specific version of the $M_h$ model (see \emph{Methods}), respectively. The two theorems we derive corresponding to the two likelihoods show that it takes miniscule quanta of data (virtually one data point in the $M_0$ case) to ensure posterior propriety when either the scale prior or the discrete uniform prior is used in conjunction with the $M_0$ and $M_h$ likelihood models, respectively. And as such, the two models do not warrant any routine application of one prior over the other. 

Indeed the cases of both the scale prior and discrete uniform prior come into the arguments when we adopt the reference prior approach to finding truly `objective' and uninformative priors \citep{berger2012objective} during Bayesian data analysis. Intuitively speaking, the reference prior theory calls for choosing priors, based on specific likelihoods using information-theoretical concepts, so that the analysis is maximally dominated by the data. Accordingly, such a basis will contradict the suggestion of a blanket application as advised in \citet{link2013cautionary}.  As such, it may be misleading to suggest that there may be a `correct' prior that ecologists should always use. 

In addition, applying such a summary rule to ecological problems may even pose some dangers in inference as seen in the applications discussed by \citet{link2013cautionary}. We discuss some of those issues here.
 
\begin{enumerate}
\item Discussing the particular relevance to data augmentation problems in ecology, \citet{link2013cautionary} warns of a major problem arising when the discrete uniform prior is chosen. In such problems, a large number of zero-inflated capture histories are included in the analysis \citep[see]{royle2008bhierarchical, royle2013spatial}. This large set, $M$, is a large support of probability mass for the inferential problem of estimating $N$, such that $N<M$. Ideally it has to be $\infty$, but owing to computational limitations it is taken to be large enough so that it is impossible to have the estimate of the parameter of interest to be larger than this value. But given that this value of $M$ set by the data analyst is of a finite value in Markov Chain Monte Carlo (MCMC) analyses, we do not see the connection between data augmentation and posterior impropriety. We do acknowledge, however, that when impropriety of the posterior exists, every unit increase in the value of $M$ should perhaps increase the value of the posterior mean in such MCMC analysis.

\item Based on mathematical results obtained in \citet{kahn1987cautionary} and \citet{york1992bayesian}, \citet{link2013cautionary} relates the practical problem on hand (analysis of snow shoe hare data) to the problems posed in \citet{kahn1987cautionary} and \citet{york1992bayesian}. In doing so, \citet{link2013cautionary} converts what is a 3-parameter problem into a 7-parameter problem using the Dirichlet's prior. The conclusion there was that posterior for $N$ was most likely improper because the conditions described in \citet{york1992bayesian} were not satisfied after conversion. This approach to determine posterior propriety is not convincing. 
 
\item We note that a blanket application of a scale prior for all situations can cause certain inferential problems. For example, if the dominating part of the probability mass on the prior distribution corresponds to the region of the maximum likelihood estimate, we should have a posterior distribution that will reduce the uncertainty to a larger extent than shown by the data. In fact, we believe that is the effect seen in the data analysis of snow shoe hares in \citet{link2013cautionary}. In the re-analysis of the snowshoe hare data, \citet{link2013cautionary} demonstrates that when the data augmentation parameter $M$ is extended from 200 to 1000, the analysis of the capture-recapture data with a scale prior had a far reduced posterior standard deviation compared to the same analysis with a discrete uniform prior. \citet{link2013cautionary} concludes that this was indicative of impropriety. This may have been the case. But we do not see evidence of it, since changes in posterior mean of $N$ was not assessed against increases in $M$. Instead, the reduction in the posterior standard deviation may be occuring due to the setting of $M$ itself because the posterior mean of $N$ (which is the parameter of interest in that problem) lies at a value of $\approx 100$ which is $0.1 \times M = 0.1 \times 1000$, and the scale prior has a large probability mass at 0.1, because the model used is $N|\psi,M \sim \mathrm{Binomial}[M,\psi]$, where $\psi$ is the probability of an individual being a member of the real population. 
\end{enumerate}

While submit that \citet{link2013cautionary} is provided only as cautionary note. However, the uptake of the recommendation of the routine use of scale prior has been quite immediate and extensive (see \emph{Introduction}) without much contest or appropriate justification, but see \citet{villa2014cautionary, link2014cautionary}. We are of the opinion that much depends on the likelihoods used to confront the data. The likelihoods used by \citet{kahn1987cautionary} and \citet{york1992bayesian} are very inefficient likelihoods and most of the problems lie here. We believe that statistical ecologists have developed far more efficient likelihoods \citep{williams2002analysis}. Consequently, the issues of impropriety should rarely be a matter of serious concern if the sources of information from data is used to carefully build the likelihoods along with well defined sampling situations . We do not look further into the issue of  `Bayesian stupefaction' discussed in \citet{link2013cautionary}. However, this oddity implies a fundamentally faulty approach of Bayesian inference, but it perhaps relates close to the efficiencies of likelihoods we have discussed in conjunction with quanta of data available for analysis. 

We suggest future research to focus on developing efficient likelihoods, and when necessary, use assessments of posterior propriety by studying the asymptotic behavior of functions as we demonstrate using the Big-$O$ notation. More broadly, however, we caution against the direct uptake of mathematical results into ecology without suitable modifications or critical assessment. Our views echoes some earlier views on the uses of mathematics in ecology \citep{may2004uses}.

\section{Acknowledgements}
We thank Indian Statistical Institute and Wildlife Conservation Society, New York for supporting this study. We thank Femke Broekhuis for asking questions about posterior propriety that motivated us to look at this problem.

\bibliographystyle{ecology}
\bibliography{ArjunBibLibrary}

\clearpage
\section{Appendices}
\subsection{Proof of Theorem for $M_0$ model.}
\noindent {\bf Theorem 1.} For the Bayesian analysis of model (\ref{cr-model}), consider the joint prior density $\pi(p,N) = \pi_1(p) \pi_2(N)$ where $\pi_1(p)$ is the density of
Beta($a$,$b$). Then the posterior distribution of $N$ is proper if and only if
$n_{.}-M_{K+1} +a > 1$ for the discrete uniform prior $\pi_2(N) \propto 1$; and it is
proper if and only if $n_{.}-M_{K+1} +a > 0$ for the scale prior, $\pi_2(N) \propto 1/N$.

\noindent {\bf Proof.} Integrating out $p$ from the joint posterior density of (\ref{cr-model}), we get
\begin{eqnarray}
\pi(N|\{y_{ij}\}) &\propto & \pi_2(N) \int_0^1 \frac{N!}{[N-M_{K+1}]!}
p^{n.}(1-p)^{KN-n.} \pi_1(p)\,dp\nonumber\\
&\propto & \pi_2(N) \frac{N!}{[N-M_{K+1}]!}
\int_0^1 p^{n.}(1-p)^{KN-n.} p^{a-1}(1-p)^{b-1}\,dp\nonumber\\
&= & \pi_2(N) \frac{N!}{[N-M_{K+1}]!}
\int_0^1 p^{n.+a-1}(1-p)^{KN-n.+b-1} \,dp\nonumber\\
&\propto & \pi_2(N) \frac{N!}{[N-M_{K+1}]!}
\frac{\Gamma(n.+a)\Gamma(KN-n.+b)}{\Gamma(KN+a+b)}\nonumber\\
&\propto & \pi_2(N) \frac{N!}{[N-M_{K+1}]!}
\frac{\Gamma(KN-n.+b)}{\Gamma(KN+a+b)}.\label{factors}
\end{eqnarray}

Note that the propriety of the posterior of $N$ depends on the asymptotic behavior of the second term (everything else but prior for $N$) in (\ref{factors}) above. We recall two results here:\\
(\romannumeral1) $\Gamma(\alpha +1) = \alpha\Gamma(\alpha)$ for any $\alpha>0$.\\
(\romannumeral2) Stirling's approximation: As $\alpha\rightarrow \infty$,
$$\frac{\sqrt{2\pi}\exp(-\alpha)\alpha^{\alpha-1/2}}{\Gamma(\alpha)}\rightarrow 1.$$

Intuitively, we would expect that the number of recaptures, say $r = n_{.}-M_{K+1}$, carries important information in the $M_0$ model. Then
\begin{eqnarray*}
\lefteqn{\frac{N!}{[N-M_{K+1}]!} \frac{\Gamma(KN-n.+b)}{\Gamma(KN+a+b)}
= \frac{N!}{[N-M_{K+1}]!} \frac{\Gamma(KN-M_{K+1}-r+b)}{\Gamma(KN+a+b)}}\\
&=& \frac{N(N-1)\cdots (N-M_{K+1}+1)}{(KN-1+a+b)(KN-2+a+b)\cdots (KN-M_{K+1}-r+a+b)}\\
&&\times \frac{\Gamma(KN-M_{K+1}-r+b)}{\Gamma(KN-M_{K+1}-r+a+b)}.
\end{eqnarray*}

Consider the first factor above. Its numerator has $M_{K+1}$ terms involving $N$.
The denominator has $M_{K+1}+r$ terms involving $KN$. This excess of $r$ terms in the denominator makes this factor asymptotically $O(N^{-r})$. Now consider the second factor. 
\begin{eqnarray*}
\lefteqn{\frac{\Gamma(KN-M_{K+1}-r+b)}{\Gamma(KN-M_{K+1}-r+a+b)}}\\
&\sim & \frac{\exp(-(KN-M_{K+1}-r+b))}{\exp(-(KN-M_{K+1}-r+a+b))}
\frac{(KN-M_{K+1}-r+b)^{KN-M_{K+1}-r+b-1/2}}{(KN-M_{K+1}-r+a+b)^{KN-M_{K+1}-r+a+b-1/2}}\\
&=&\exp(a) \left(\frac{KN-M_{K+1}-r+b}{KN-M_{K+1}-r+a+b}\right)^{KN-M_{K+1}-r+b-1/2}
(KN-M_{K+1}-r+a+b)^{-a}\\
&=& \exp(a) \left(1-\frac{a}{KN-M_{K+1}-r+b-1/2}\frac{KN-M_{K+1}-r+b-1/2}{KN-M_{K+1}-r+a+b}\right)^{KN-M_{K+1}-r+b-1/2}\\
&& \times (KN-M_{K+1}-r+a+b)^{-a}\\
&\sim & \exp(a) \exp(-a) O(N^{-a}).
\end{eqnarray*}
Thus we have that
\begin{eqnarray}
\pi(N|\{y_{ij}\}) &= & \pi_2(N) O(N^{-(r+a)})\nonumber\\
&= & \pi_2(N) O(N^{-(n_{.}-M_{K+1}+a)}).\label{M_0proof}
\end{eqnarray}

Therefore, from (\ref{M_0proof}), the determination of propriety will depend on the product of the prior $\pi_2(N)$ and Big-$O$ evaluation of the likelihood, $O\left(N^{-(n_{.}-M_{K+1}+a)}\right)$. If $d=n_{.}-M_{K+1}+a$, then for a discrete uniform prior ($\propto 1$), we will need $c > 1$ for propriety and for a scale prior ($\propto 1/N$), we will need $d > 0$ for propriety, and thus the proof.

\subsection{Proof for Theorem for $M_h$ model.}

\noindent {\bf Theorem 2.} For the Bayesian analysis of model (\ref{int-mh-cr-model}),
 consider the joint prior density $\pi(\alpha,\beta ,N) = \pi_1(\alpha,\beta) \pi_2(N)$ where,
under $\pi_1$, $\alpha$ and $\beta$ are independent Gamma random variables with shape
parameters $a>0$ and $b>0$ respectively and a common scale parameter $c$. Then the posterior distribution of $N$ is proper if $a > 1$ for the discrete uniform prior $\pi_2(N) \propto 1$; and it is 
proper if $a > 0$ for the scale prior, $\pi_2(N) \propto 1/N$. 

\noindent {\bf Proof.} Since
\begin{eqnarray*}
\lefteqn{\pi(N,\alpha,\beta|\{y_{ij}\})}\\
&\propto & \pi_2(N) \pi_1(\alpha,\beta) P(\{y_{ij}\}|N,\alpha,\beta) \\
&\propto & \pi_2(N) \pi_1(\alpha,\beta) \frac{N!}{[M_{K+1}]! [N-M_{K+1}]!}
\left(\frac{\prod_{j=0}^{K-1}(\beta+j)}{\prod_{j=0}^{K-1}(\alpha+\beta+j)}\right)^{N-M_{K+1}}\\
&&\times \prod_{i=1}^{M_{K+1}}\left(\frac{\prod_{j=0}^{y_{i.}-1}(\alpha+j)\prod_{j=0}^{K-y_{i.}-1}(\beta+j)}
{\prod_{j=0}^{K-1}(\alpha+\beta+j)}\right),
\end{eqnarray*}
we have that
\begin{eqnarray*}
\lefteqn{\pi(N|\{y_{ij}\})=\int \pi(N,\alpha,\beta|\{y_{ij}\})\,d\alpha\,d\beta}\\
&\propto & \pi_2(N) \int_0^\infty\int_0^\infty \pi_1(\alpha,\beta) P(\{y_{ij}\}|N,\alpha,\beta)\,d\alpha\,d\beta\\
&\propto & \pi_2(N) \frac{N!}{[N-M_{K+1}]!}
E\bigl[\left(\frac{\prod_{j=0}^{K-1}(\beta+j)}{\prod_{j=0}^{K-1}(\alpha+\beta+j)}\right)^{N-M_{K+1}}\\
&&\times \prod_{i=1}^{M_{K+1}}\left(\frac{\prod_{j=0}^{y_{i.}-1}(\alpha+j)\prod_{j=0}^{K-y_{i.}-1}(\beta+j)}
{\prod_{j=0}^{K-1}(\alpha+\beta+j)}\right)\bigr],
\end{eqnarray*}
where the expectation above is with respect to the joint prior distribution $\pi_1$ of $(\alpha,\beta)$.
Once again, as we investigated the propriety conditions with respect to $M_0$ model, it is sufficient to investigate the \emph{asymptotic behavior} of this expectation (relative to the other
factors involving $N$ in the posterior density) as $N\rightarrow \infty$
to determine the propriety of the posterior distribution of $N$. For this, we note that
\begin{eqnarray*}
\left(\prod_{j=0}^{K-1}\frac{\beta+j}{\alpha+\beta+j}\right)^{N-M_{K+1}}
&=& \left(\frac{\beta}{\alpha+\beta}\right)^{N-M_{K+1}}
\left(\prod_{j=1}^{K-1}\frac{\beta+j}{\alpha+\beta+j}\right)^{N-M_{K+1}}\\
&\le & \left(\frac{\beta}{\alpha+\beta}\right)^{N-M_{K+1}},
\end{eqnarray*}
and also
\begin{eqnarray*}
\lefteqn{\prod_{i=1}^{M_{K+1}}\left(\frac{\prod_{j=0}^{y_{i.}-1}(\alpha+j)
\prod_{j=0}^{K-y_{i.}-1}(\beta+j)} {\prod_{j=0}^{K-1}(\alpha+\beta+j)}\right)}\\
&=& \prod_{i=1}^{M_{K+1}}\left(\prod_{j=0}^{y_{i.}-1}\frac{\alpha+j}{\alpha+\beta+j}
\prod_{j=0}^{K-y_{i.}-1}\frac{\beta+K-y_{i.}-1-j}{\alpha+\beta+K-1-j}\right)\\
&=& \left(\prod_{i=1}^{M_{K+1}} \frac{\alpha}{\alpha+\beta}\right)
\prod_{i=1}^{M_{K+1}}\left(\prod_{j=1}^{y_{i.}-1}\frac{\alpha+j}{\alpha+\beta+j}
\prod_{j=0}^{K-y_{i.}-1}\frac{\beta+K-y_{i.}-1-j}{\alpha+\beta+K-1-j}\right)\\
&\le & \left(\frac{\alpha}{\alpha+\beta}\right)^{M_{K+1}}.
\end{eqnarray*}
Therefore,
\begin{eqnarray*}
\lefteqn{E\bigl[\left(\frac{\prod_{j=0}^{K-1}(\beta+j)}{\prod_{j=0}^{K-1}(\alpha+\beta+j)}\right)^{N-M_{K+1}}
 \prod_{i=1}^{M_{K+1}}\left(\frac{\prod_{j=0}^{y_{i.}-1}(\alpha+j)\prod_{j=0}^{K-y_{i.}-1}(\beta+j)}
{\prod_{j=0}^{K-1}(\alpha+\beta+j)}\right)\bigr]}\\
&\le & E\bigl[\left(\frac{\beta}{\alpha+\beta}\right)^{N-M_{K+1}}
\left(\frac{\alpha}{\alpha+\beta}\right)^{M_{K+1}}\bigr]\\
&=& E\bigl[(1-X)^{N-M_{K+1}} X^{M_{K+1}}\bigr],
\end{eqnarray*}
where $X = \alpha/(\alpha+\beta) \sim \mbox{ Beta}(a,b)$.
Hence,
\begin{eqnarray*}
E\bigl[\left(\frac{\beta}{\alpha+\beta}\right)^{N-M_{K+1}}
\left(\frac{\alpha}{\alpha+\beta}\right)^{M_{K+1}}\bigr]
& =& \frac{\Gamma(a+b)}{\Gamma(a)\Gamma(b)}
\frac{\Gamma(M_{K+1}+a)\Gamma(N-M_{K+1}+b)}{\Gamma(N+a+b)}\\
&=& O(N^{-(M_{K+1}+a)})
\end{eqnarray*}
using an argument similar to that in the proof of Theorem~1. Therefore,
\begin{eqnarray*}
\pi(N|\{y_{ij}\})
&\propto & \pi_2(N) \frac{N!}{[N-M_{K+1}]!}
O(N^{-(M_{K+1}+a)})\\
&=& \pi_2(N) O(N^{-a}).
\end{eqnarray*}

\end{spacing}
\end{document}